\documentclass[aps,prl,twocolumn,10pt,colorlinks,groupaddress,showpacs]{revtex4-1}
\usepackage{graphicx}
\usepackage{amssymb}
\usepackage{wasysym}
\usepackage{natbib}
   \usepackage[dvipdfmx,colorlinks=true,citecolor=blue,linkcolor=black,pdfhighlight =/O]{hyperref} % si compilation dvi ou ps (PR)
\usepackage{color}
\usepackage{soul}

\begin{document}
%\ifx\pdfoutput\undefined
%   not running PDFTeX
%\else
%   running PDFTeX
%\fi
% Use the \preprint command to place your local institutional report number in the upper righthand corner of the title page in preprint mode.
%\preprint{}
\title{Sub-Kelvin tunneling spectroscopy showing Bardeen-Cooper-Schrieffer superconductivity in heavily boron-doped silicon epilayers}

\author{F.~Dahlem$^{1}$, T.~Kociniewski$^{2}$, C.~Marcenat$^{3}$, A.~Grockowiak$^{1,3}$, L.M.A.~Pascal$^{1}$, P.~Achatz$^{1}$, J.~Boulmer$^{2}$, D.~D\'ebarre$^{2}$, T.~Klein$^{1}$, E.~Bustarret$^{1}$, H. Courtois$^{1}$}
%\email[Email address: ]{franck.dahlem@grenoble.cnrs.fr}

\affiliation{
$^{1}$ Institut N\'eel, CNRS and Universit\'e Joseph Fourier, Grenoble, France.\\
$^{2}$ CNRS and Universit\'e Paris Sud, Institut d'Electronique Fondamentale, Orsay, France.\\
$^{3}$ CEA-Grenoble, Institut Nanosciences et Cryog\'enie, SPSMS-LATEQS, Grenoble, France.
}
\date{\today}
\begin{abstract}
Scanning tunneling spectroscopies in the subKelvin temperature range were performed on superconducting silicon epilayers doped with boron in the atomic percent range. The resulting local differential conductance behaved as expected for a homogeneous superconductor, with an energy gap dispersion below $\pm 10\%$. The spectral shape, the amplitude and temperature dependence of the superconductivity gap follow the BCS model, bringing support to the hypothesis of a hole pairing mechanism mediated by phonons in the weak coupling limit.
\end{abstract}

\maketitle

Upon heavy boron doping, silicon shows a superconductive resistive and diamagnetic transition at low temperature~\cite{bustarret465}. This recent discovery has been held as a significant milestone~\cite{cava427} because of the archetypical nature of this covalent cubic semiconductor~\cite{Blase375,Iakoubovskii2009675,ren103710,Herrmannsdorfer217003}. Similarly to diamond, the charge carriers involved in silicon superconductivity are generally thought to be holes at the Fermi level lying within and near the top of the valence band~\cite{bourgeois142511,blase237004}. Unlike diamond~\cite{bustarret237005}, the superconductive transition in silicon requires a boron content (several percents~\cite{marcenat020501}) much higher than the one needed for the insulator to metal transition (about 80~ppm~\cite{dai1914}). This threshold also lies above the boron solubility limit in crystalline silicon, raising the question of the homogeneity of doping and hence superconductivity in Si:B epilayers obtained by out-of-equilibrium growth. On a more fundamental basis, the nature of the pairing mechanism in this superconductor remains to be determined. So far, the critical temperature values, the optical phonon softening observed by Raman spectroscopy in the normal state~\cite{bustarret465} and the ab initio calculations of the electron-phonon coupling~\cite{bourgeois142511} have merely shown that the overall features of this superconductor were compatible with a conventional pairing mechanism. These open questions call for local investigations of superconductivity in doped silicon.  

In this Rapid Communication, we report a detailed study in heavily boron-doped silicon by subKelvin tunneling spectroscopy. The local electronic density of states (LDOS) spectral shape, the amplitude and temperature dependence of the related energy gap in the excitation spectrum are consistent with the picture of a superconductor following the BCS model. The superconducting energy gap was found to vary locally by up to $\pm 10\%$.  

\begin{figure}[b]
\includegraphics[width=\columnwidth]{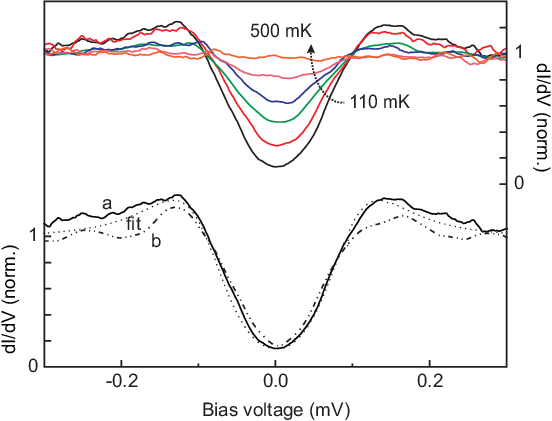}
\caption{(Color online)
Bottom: normalized differential conductances ($dI/dV$) of tunnel contacts on superconducting silicon (sample B1, \mbox{$\rm{T_{sample}}=110$~mK}) at one micron apart positions: a and b. The dotted line is a BCS fit.
Top: variation of the $dI/dV$ spectrum with respect to sample temperature from bottom to top $\rm{T_{sample}}$ = 110, 350, 425, 440, 460, 500~mK on the same sample and at position $a$. The set-point tunnel resistance was 1~M$\Omega$. 
\label{fig1_dIdV}}
\end{figure}

Superconductive silicon samples were prepared by gas immersion laser doping (GILD)~\cite{Kerrien200245,*Kerrien2003277}. A precursor gas atmosphere BCl$_3$ was injected on a $<$100$>$-oriented silicon wafer surface in order to saturate silicon chemisorption sites by boron atoms. After each injection, the surface was melted by an ultraviolet laser pulse, which inserted boron atoms by diffusion into the silicon wafer. This out-of-equilibrium method allows substitutional boron atoms to be incorporated into the silicon lattice well beyond the thermodynamic solubility limit of about a few  $\rm{10^{20}~cm^{-3}}$ (\mbox{i.e.} below one atomic percent). The doped area determined by the laser spot size was typically 2 by 2 square millimeters. As a result, high concentration secondary ion mass spectroscopy (SIMS) profiles~\cite{marcenat020501, dubois1377} show, below a few nanometer-thick surface oxide, an homogeneously doped layer thickness extending over 30 to 130 nanometers depending on the samples preparation.
 
In the present work, we studied different samples (A1, B\{1,2,3\} and C1), which were prepared by respectively 200, 500 and 400 laser shots, yielding  different 'macroscopic' critical temperatures $T_c$= 350, 510 and 590~mK. The above-mentioned values of the critical temperature $T_c$ were determined as the temperature corresponding to a resistance below 10~\% of the normal state value. This criterion was chosen because this critical temperature coincides with that of the magnetic susceptibility onset~\cite{marcenat020501}.   

In sample A1, which corresponds to the first set of superconducting silicon studied in Ref.~\cite{bustarret465}, the laser melting duration was short (25~ns) and the layer was very thin (30~nm), giving a low critical temperature $T_c$ value and a large transition width of $\Delta T \approx$ 80~mK. This width is defined from 10\% to 90\% of the resistive transition. By increasing the number of laser shots, it was possible to increase the layer thickness (80~nm) and optimize the critical temperature~\cite{marcenat020501}. For instance, sample C1 has the highest critical temperature $T_c$= 590~mK and the sharpest transition: $\Delta T \approx$~20~mK. In this sample, a mean free path $l$ of 3~nm and a superconducting coherence length $\xi_S$ of about $50$~nm have been previously estimated via the square resistance $R_{\,\square} \approx 17~\Omega$ at room temperature and the upper critical magnetic field, respectively~\cite{marcenat020501}. In the case of a large number of laser shots, a fraction of boron atoms is not incorporated on substitutional sites~\cite{marcenat020501}. For samples B, which exhibit a 130~nm thickness, this leads to a more disordered silicon structure, a lower critical temperature $T_c$ value and a larger transition width: $\Delta T \approx$~40~mK.   

From the critical temperature values discussed above, the energy of the local superconducting gap is expected to be in the low milli-electronvolt range. The use of our home-built scanning tunneling microscope (STM) working in a dilution refrigerator~\cite{Moussy-RSI} is therefore crucial to achieve a local spectroscopy with the proper energy resolution. This surface probe has been for instance applied to superconducting  diamond~\cite{dahlem09123727,sacepe097006} and it is particularly well suited to study thin films such as Si:B epilayers. Furthermore, STM probing has the advantage to give access to the intrinsic sample inhomogeneity.

For local tunneling spectroscopy, a sweep of bias voltage $V$ was applied between the silicon sample and a metallic tip~\footnote{We have mainly used etched tungsten (W) or platinum-iridium (PtIr) tips. A few spectroscopies were also performed with niobium (Nb) superconducting tips.} at a fixed sample-tip distance, while the induced tunnel current $I$ was recorded. We used two types of measurement that gave similar results: an \mbox{a.c.} lock-in technique with a $\rm{15~\mu V}$ excitation at about $\rm{2~kHz}$ frequency and a numerical derivation from a \mbox{d.c.} $I(V)$ measurement. At thermal equilibrium, the local differential conductance ($dI/dV$) gives access to the local density of states $\rho$ smeared out by the thermal energy $\rm{3.5\,k_{B} T}$: $dI/dV \propto \int_{\Re}\rm{dE\,\partial f / \partial E \cdot \rho(E-eV)}$. In this expression, the energy distribution function derivative with respect to the energy $\rm{\partial f / \partial E}$ features the thermal broadening. For very low temperature experiments, we define an effective temperature in order to include several additional effects, which also smear out the electronic current characteristic. These contributions are related to local heat induced by out of equilibrium electron injection, electromagnetic noise (black-body radiation of hot objects, imperfect filtering), weak electron-phonon coupling. The effective temperature is usually higher than the refrigerator base temperature and it determines the practical energy resolution of the measurement. In our case, this resolution is better than  $\rm{2\,k_{B} T_{eff}(300~mK) \approx 50~\mu eV}$.

\begin{figure}[b]
\includegraphics[width=\columnwidth]{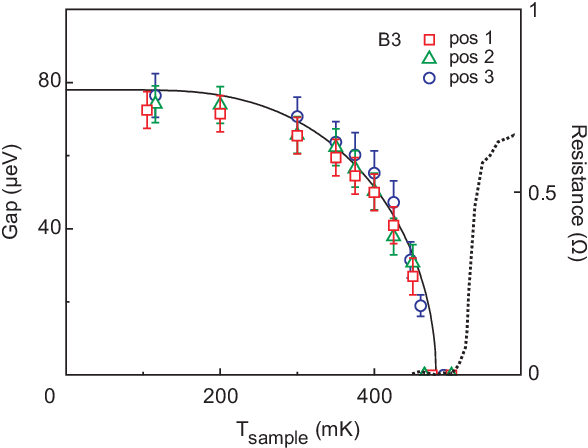}
\caption{(Color online)
Left axis: variation of the superconducting energy gap versus the sample temperature, recorded for three different positions in sample B3. 
Position 1 and 2 are 15~nm apart, whereas position 3 is 100~nm away. The continuous line is to a fit to the BCS theory. 
Right axis: electrical resistance of the same sample versus temperature (dotted line), showing a macroscopic superconductive transition with a critical temperature (see text for its precise definition) \mbox{$T_c \approx510$~mK}. 
\label{fig2_GapT}}
\end{figure}

Fig.~\ref{fig1_dIdV} bottom presents local differential tunnel conductances measured at a sample temperature \mbox{$T_{\rm{sample}}= 110$~mK} on sample B1. A superconducting energy gap is clearly visible at low bias voltage, with two coherence peaks around $\left|V\right|=0.13$~mV. A constant differential conductance characteristic of an ohmic behavior is recovered at higher voltage. Such a spectroscopy measurement represents the local evidence of superconductivity in heavily boron-doped silicon. A series of positions were probed. Fig.~\ref{fig1_dIdV} displays as an illustrative example spectra acquired at two positions $a$ and $b$ distant from one micrometer, which is one order of magnitude larger than the coherence length. Similar spectra were found in all cases, which indicates that the superconductivity is spread over the whole sample with a similar behavior. The signature of inelastic tunneling processes involving softened optical phonons at 450 and 600~$cm^{-1}$~\cite{bustarret465} (i.e. 55-75 meV) has also been sought in the second derivative $d^{2}I/dV^{2}$ of the tunnel current~\cite{solymar_book}, but without success.

The differential conductance $dI/dV$ spectrum can be fitted by a BCS-type density of states (dotted line in Fig.~\ref{fig1_dIdV}). This expression does not include any Dynes parameter but takes into account a fixed effective electronic temperature of 300~mK. A gap amplitude about $\Delta(T_{\rm{sample}} = 110~\rm{mK}) = 87\pm7\,\rm{\mu eV}$ is extracted from the fit. The experimental error bar in the gap value is significantly smaller than could be naively expected from the thermal broadening, since the fit is obtained with the gap as the single adjustable parameter. 

Recent tunnel junction experiments performed on specific arrays of microdefects in doped silicon have yielded a large energy gap, which was attributed to a high-temperature superconductivity up to 150 Kelvin~\cite{bagraev21,*bagraev0806}. Here, in order to confirm the superconductivity origin of the measured gap of excitations, we varied the sample temperature and recorded the corresponding local differential conductance. A typical set of differential conductance data at different temperatures, obtained for sample B1 at a single position $a$, is presented at the top of Fig.~\ref{fig1_dIdV}. As expected, the amplitude of the energy gap decreases by increasing the sample temperature (lower to upper curve at zero bias), while the smearing also increases. 

At every sample temperature, the value of the superconductivity gap can be obtained by fitting the spectrum with a BCS expression, which allowed us to determine its temperature dependence. Fig.~\ref{fig2_GapT} presents this data for three different positions (1, 2 and 3) in sample~B3. The distance between 1 and 2 on one hand and (1,2) and 3 on the other hand was respectively below and above the estimated superconducting coherence length. At these three different positions, the energy gap values coincide within the error bars. 

The temperature dependence of the gap is also well fitted by the thermally-smeared BCS prediction displayed as the continuous line in Fig.~\ref{fig2_GapT}. It follows for sample~B3 a zero temperature energy gap $\rm{\Delta(0) = 78\pm 9~\mu eV}$ and a local critical temperature $T_c =$~480~$\pm$ 30~mK. The critical temperature accessed in this way is a local quantity. It corresponds to a value close to the tail of the superconductive resistive bulk transition (see dotted curve in Fig.~\ref{fig2_GapT}), which sits at $T_c \approx$~510~mK. This distinction between the 'resistive' and the local critical temperature properly reflects the difference between the transport measurement and the local probing. 
 
In total, we have measured five samples, and more than one hundred spectra have been obtained on different spatial positions. All curves showed similar results, demonstrating that the data displayed in the previous Figures are characteristic of superconductive heavily boron-doped silicon. The quality of the tunnel junction has been controlled by recording the differential conductance $dI/dV$ at different tunneling resistances, \mbox{i.e.} different sample-tip distances. When changing the set-point tunnel resistance between $\rm{10~M\Omega}$ and $\rm{1~M\Omega}$, the tunneling spectra overlap after renormalization with respect to the normal differential conductance, as expected. Since samples stayed under air atmosphere before being transferred inside the cryogenics microscope, buffered oxide etch (BOE) or diluted hydrofluoric acid (HF) passivating has sometimes been performed to warrant a clean surface, but no significant difference has been revealed compared to the 'as grown' surface. Furthermore, we succeeded in fabricating a millimeter square solid superconductive-normal metal tunnel junction to probe the tunneling current averaged on a large scale. The measured current-voltage characteristic provided a similar differential conductance spectrum with a superconducting bulk gap value comparable to the one measured locally by STM. 

\begin{figure}[b]
\includegraphics[width=\columnwidth]{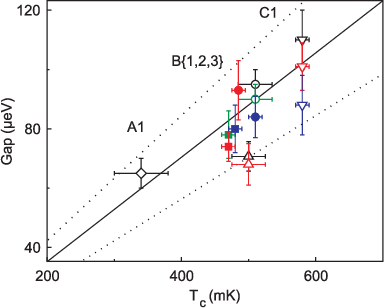}
\caption{(Color online)
Dependence of the extrapolated zero-temperature superconducting gap $\Delta (0)$ with the critical temperature $T_c$ for series A, B and C samples. 
For full symbols, $T_c$ was locally determined through the local energy gap temperature dependence.  
For empty symbols, $T_c$ was obtained via macroscopic transport measurements. 
The straight line indicates the BCS ratio $\Delta (0)/k_BT_c = 1.76$, while the dotted lines show a $\pm~20$\% deviation. 
Sample B1 and C1 data with empty ($\Circle$, $\triangledown$) and full ($\CIRCLE$) symbols are measured at a macroscopic distance (\mbox{i.e.} at different runs). 
Sample B2 data ($\triangle$) were measured $\rm{0.5~\mu m}$ apart. 
Sample B3 symbols ($\blacksquare$) summarize Fig.~\ref{fig2_GapT} results. 
\label{fig3_GapTc}}
\end{figure}

Finally, the difference in critical temperature of the three series A, B and C enabled us to investigate the evolution of the zero temperature superconducting energy gap value with the critical temperature. Fig.~\ref{fig3_GapTc} displays the series of ($\Delta (0),\, T_c$) couples for the whole experimental runs set on the five investigated samples. The critical temperatures $T_c$ were determined either by macroscopic transport measurement (empty symbols in Fig.~\ref{fig3_GapTc}) or, when available, by local temperature dependence measurements, following a similar procedure such as the one shown in Fig.~\ref{fig2_GapT} (full symbols in Fig.~\ref{fig3_GapTc}). In the latter case, the local critical temperature was always lower than the macroscopic value. All measured data points are located around the full line corresponding to the BCS model prediction $\Delta (0)/k_B\,T_c$=1.76. Data with a critical temperature measured locally mainly agree with the BCS ratio within the error bars. For the rest of the data, gap values smaller than the BCS prediction are ascribed to a weakening of superconducting energy gap, which may occur locally because of a higher disorder or a lower charge doping at the surface compared to the bulk of the film.
  
On a given sample, the measured energy gaps show a maximum $\pm10~\%$ scattering. The same scattering is observed over small and large distances, see for instance sample B1 data in Fig. 3. The disorder accessed through the local superconductive energy gap thus sits at a very small scale, of the order of the mean free path. The energy gap variation normalized to the mean value is larger than the width of the resistive transition normalized to the critical temperature. Actually, the local energy gap dispersion measured by STM gives an accurate picture of the actual spatial variation of the superconductivity, whereas the resistive transition indicates only the appearance of a superconducting percolating path throughout the sample. These two approaches can give a different picture, depending on the investigated sample. For instance, sample C1 yielded the sharpest superconducting resistive transition~\cite{marcenat020501} of all samples but an energy gap dispersion similar to the one observed in series B sample with a broader resistive transition. Finally, let us point out that although the gap inhomogeneity observed here is reminiscent of predictions on superconductivity in highly disordered metals close to the metallic to insulator transition~\cite{feigelman027001}, our samples have a doping level two orders of magnitude higher than the relevant critical threshold. 

In conclusion, the tunnel differential conductance spectral shape, the BCS-compatible dependence of the superconducting gap with the temperature, as well as the $\Delta (0)/k_B\,T_c$ ratio close to the BCS value, are three strong experimental indications that superconductivity in heavily boron-doped silicon follows the weak coupling BCS model. The microscopic gap variation of a few percents fits well with expectations for a strongly disordered metal like highly boron-doped silicon. This progress in understanding superconductivity in silicon opens the way to the future completion of reliable monolithic devices based on hybrid superconductor/normal junctions with heavily boron-doped regions as the superconductor.
 
The authors thank C.~Dubois, G.~Prudon, J.-C.~Dupuy and B.~Gautier for SIMS measurements, R.~Sukumar and Th.~Quaglio for discussions. 
We acknowledge the support from MICROKELVIN, the EU FRP7 low temperature infrastructure grant 228464, R\'egion Rh\^one-Alpes and ANR contract ANR-08-BLA-0170.

%\bibliography{dahlem_susiSTM}

%Merlin.mbs v4.21 2009-07-09.
%

\end{document}